# Reversible quantum Brownian heat engines for electrons

T.E. Humphrey [1], R. Newbury [1], R.P. Taylor [2], and H. Linke [2,*]

[1] School of Physics, University of New South Wales, UNSW Sydney 2052, Australia
[2] Department of Physics, University of Oregon, Eugene OR 97403-1274, U.S.A.

* Corresponding author.
Email: linke@darkwing.uoregon.edu

Brownian heat engines use local temperature gradients in asymmetric potentials to move particles against an external force. The energy efficiency of such machines is generally limited by irreversible heat flow carried by particles that make contact with different heat baths. Here we show that, by using a suitably chosen energy filter, electrons can be transferred *reversibly* between reservoirs that have different temperatures and electrochemical potentials. We apply this result to propose heat engines based on mesoscopic semiconductor ratchets, which can quasistatically operate arbitrarily close to Carnot efficiency.

Ratchets combine asymmetry with non-equilibrium processes to generate directed particle motion [1]. When non-equilibrium is induced by contact with heat baths at different temperatures, a ratchet can act as a so-called Brownian heat engine, converting local spatial or temporal temperature variations into useful work [2, 3]. This mechanism is currently attracting considerable interest [4-6] and in the future may be applied to power artificial micro-machines such as chemical motors [5, 7]. Many proposed Brownian heat engines, however, have a maximum theoretical efficiency that is significantly lower than the ideal Carnot efficiency. This critical limitation to potential applications can generally be traced to irreversible heat flow via a degree of freedom of the ratchet mechanism itself, namely the kinetic energy of particles making contact with different heat baths [4, 5].

Derenyi and Astumian analysed overdamped Brownian particles moving between different heat baths and found that Carnot efficiency *can* be achieved in the quasistatic limit where the kinetic energy of particles approaches zero [5]. Here, we present a novel mechanism for reversible particle transfer between heat baths that is suitable for moderately damped particles with arbitrary kinetic energy. We consider an electronic system that allows independent control of gradients in temperature and in electrochemical potential. Our main result is that two electron reservoirs with different temperatures and electrochemical potentials can exchange electrons reversibly at the energy where the Fermi-Dirac distributions in the two reservoirs have the same value. Using suitable energy filters, one can construct Brownian heat engines that in principle can attain Carnot efficiency.

We begin our analysis by considering the transfer of single electrons between two electron reservoirs, denoted $L$ and $R$ (Fig. 1a). Each reservoir is described by an equilibrium Fermi-Dirac distribution with temperatures $T_L$ and $T_R$ and electrochemical potentials $\mu_L$ and $\mu_R$, respectively. A bias voltage $V$ applied to reservoir $R$ creates a difference in the electrochemical potentials, $(\mu_R - \mu_L) = -eV$, where $(-e)$ is the charge on the electron. Except for the flow of electrons between reservoirs, the reservoirs are assumed to be thermally isolated. The electron mean free path for inelastic processes is taken to be much larger than the distance between the reservoirs, but much smaller than the reservoir's dimension (so-called moderate damping). This regime of transport is found in mesoscopic semiconductor devices at temperatures of a few Kelvin and below [8].

The heat associated with the addition of one electron to a reservoir (electrochemical potential $\mu$) is given by $\Delta Q = \Delta U - \mu$, wherein $\mu$ is assumed to remain unchanged, and where $\Delta U$, the increase in internal energy, is given by the electron energy. As an example, we consider an electron traveling from reservoir $L$ to reservoir $R$ at constant energy $\varepsilon$, where $\varepsilon$ is measured relative to the same voltage-independent zero as the electrochemical potentials. The change of heat in $L$ associated with the electron transfer is then $\Delta Q_L = -(\varepsilon - \mu_L)$, where the negative sign in front of the bracket indicates removal of heat. The heat added to $R$ is $\Delta Q_R = (\varepsilon - \mu_R)$ and exceeds      $-\Delta Q_L$ by $eV$, because the electron picks up kinetic energy in the electric field between reservoirs.

The overall increase in entropy due to the transfer of one electron from $L$ to $R$ is

$$\Delta S = \frac{\Delta Q_L}{T_L} + \frac{\Delta Q_R}{T_R} = \frac{-(\varepsilon - \mu_L)}{T_L} + \frac{(\varepsilon - \mu_R)}{T_R}. \qquad (1)$$

Figure 1 illustrates the meaning of this equation. Consider two reservoirs at the same temperature $T$, but with different electrochemical potentials, $(\mu_L - \mu_R) = eV > 0$ [Fig. 1a]. Independent of the electron energy, the transfer of an electron from $L$ to $R$ increases the system entropy by $\Delta S = eV/T > 0$. In the complementary case, when $T_R > T_L$, but $V = 0$ (see Fig. 1b), $\Delta S = [\varepsilon - \mu] (1/T_L - 1/T_R)$. This means it is thermodynamically advantageous ($\Delta S > 0$) if, on average, 'cold' electrons ($\varepsilon < \mu$) move from $L$ to $R$ and, by symmetry, 'warm' electrons ($\varepsilon > \mu$) move from $R$ to $L$. If the probability for electron transmission across the junction is independent of $\varepsilon$, no net electric current is generated, but heat flows from $R$ to $L$. The increase in total entropy in both examples, Figs. 1(a) and (b), shows that these processes are spontaneous and irreversible.

Consider now the general case where $T_R \neq T_L$ and $\mu_L \neq \mu_R$ [Fig. 1c]. This is the situation encountered in Brownian heat engines - when particles driven by a temperature gradient do work against an external force. Significantly, Eq. (1) yields $\Delta S = 0$ for $\varepsilon = \varepsilon_S$, where

$$\varepsilon_S \equiv \frac{\mu_L T_R - \mu_R T_L}{T_R - T_L}. \qquad (2)$$

This is the main result of the present work: two electron reservoirs with arbitrary temperatures and electrochemical potentials can exchange electrons at energy $\varepsilon_S$ *reversibly* - that is, without increase of the system entropy.

It is instructive to understand how $\varepsilon_S$ relates to the energy distributions in the two reservoirs. In particular, the energy $\varepsilon_S$ fulfills the condition $f_L(\varepsilon_S) = f_R(\varepsilon_S)$, where $f_{R,L}(\varepsilon) = 1/(1+\exp[(\varepsilon - \mu_{R,L})/kT_{R,L}])$ are the Fermi-Dirac distribution functions in $R$ and $L$, respectively, shown in Fig. 1(d) for $T_R > T_L$ and $V > 0$. For $\varepsilon > \varepsilon_S$, the probability of finding an electron in reservoir $R$ is higher than in $L$. Electrons can thus increase the system entropy by moving from $R$ to $L$, following the temperature gradient. Electrons in the range $\varepsilon < \varepsilon_S$ can increase entropy by following the electrochemical potential gradient from L to R. For $\varepsilon = \varepsilon_s$, where the probability for finding an electron is the same on both sides, the two driving forces cancel. One may say that at this particular energy the two reservoirs behave as if they were in thermal equilibrium with each other. If the two reservoirs were connected via an ideal energy filter that was transparent for electrons at $\varepsilon_s$ and at no other energy, no time-averaged particle or heat current would occur spontaneously. The warm bath would not cool, and the voltage would drive no current!

Note that the above situation is distinct from an open-circuit thermovoltage generated across a quantum point contact between electron reservoirs where $T_R \neq T_L$ [9]. In that case, even though the net electric current is zero, electrons transported above and below $\varepsilon_s$ all increase the system entropy, so that energy needs to be expended to maintain the temperature difference.

We will now apply Eq. (2) to propose an electron ratchet that can be operated as a heat engine and we will demonstrate analytically how Carnot efficiency can be achieved using ideal energy filters. Specifically, we will consider a 'rocked' electron ratchet - that is, electrons in an asymmetric potential that is tilted periodically and symmetrically by an external force [1]. An adiabatically rocked electron ratchet is essentially a non-linear rectifier: the magnitude and spectral composition of the current depend, for asymmetric devices, on the voltage sign. Consequently, a symmetric AC 'rocking' voltage generates, on time-averaging, a net electric current [10] or heat current [11].

Fig. 2(a,b) illustrates a hypothetical non-linear device connecting two equal two-dimensional (2D) reservoirs, L and R, rocked by a square-wave voltage of amplitude $V_0$ > 0. We assume that switching between the values $\mp V_0$ occurs on a time scale slower than any characteristic electronic times, such as energy relaxation times (so-called adiabatic rocking), but much faster than the rocking period. It is therefore sufficient to analyse the device for the two DC situations $V = \mp V_0$, while transient behaviour can be neglected. The probability for electrons to be transmitted across the device is taken as a single Lorentzian resonance,

$$t(\varepsilon, \mp V_0) = t_0 \left/ \left[1 + \left\{ g \left/ \left[(\varepsilon - \varepsilon_{res})/g \right] \right. \right\}^2 \right] \right. \qquad t_0 \le 1$$

with amplitude $t_0 \le 1$ and a full width at half maximum of $2g$. The filter resonances, $\varepsilon_{res}(\mp V_0) = [\mu_0(\mp V_0) \pm \phi]$ for $V = \mp V_0$, respectively, are symmetrically arranged around $\mu_0 = 0.5[\mu_L + \mu_R(\mp V_0)]$ (see Fig. 2(a,b)). As can be confirmed using Eq. (3) below, this ensures that the time-averaged electric current $I^{net}(V_0) = 0.5[I(V_0) + I(-V_0)]$ is zero, thus avoiding the trivial condition where a finite $I^{net}$ is accompanied by a heat current. As a realization of a filter with this transmission function, one can consider coherent resonant tunneling [8] via an asymmetric quantum dot, in which the resonant energy level shifts its value when a bias voltage deforms the band structure, and which is connected by 1D quantum point contacts to 2D electron gas (2DEG) reservoirs.

The steady-state electric current from $L$ to $R$ generated by a DC bias voltage $V$ applied to $R$ is, for $(kT, |eV|) \ll \mu_0$, given by a Landauer equation [8],

$$I(V) = -\frac{2e}{h} \int_0^\infty t(\varepsilon, V)[f_L(\varepsilon) - f_R(\varepsilon, V)] d\varepsilon. \qquad (3)$$

The DC heat current entering each reservoir at the given bias voltage $V$ can be obtained from Eq. (3) by replacing the electron charge, $(-e)$, by the heat changes in $R$ and $L$ associated with each electron moving from $L$ to $R$, $\Delta Q_{R/L} = \mp(\varepsilon - \mu_{R/L})$:

$$q_{R/L}(V) = \pm \frac{2}{h} \int_0^\infty (\varepsilon - \mu_{R/L}) t(\varepsilon, V)[f_L(\varepsilon) - f_R(\varepsilon, V)] d\varepsilon. \qquad (4)$$

(The upper sign in $\pm$ refers to $R$, and the lower sign refers to $L$.) Note that the heat current is not conserved. Specifically, $q_L(V) = -q_R(V) + I(V)V$, where $I(V)/V$ is the externally supplied Joule heating power. Finally, the net heat current flowing into each of the two reservoirs, averaged over a full cycle of square-wave rocking, is given by

$$q_{R/L}^{net}(V_0) = 0.5[q_{R/L}(V_0) + q_{R/L}(-V_0)].$$

The efficiency of a heat engine, $\eta_E$, is given by the ratio of the work output to the heat removed from the warmer of the two reservoirs [6]. Assuming that $T_R > T_L$, we can write $\eta_E = (-W)/(-q_R^{net})$, where $W = (q_L^{net} + q_R^{net})$ is the electrical power input into the device, averaged over a full cycle of rocking. The coefficient of performance of a refrigerator, cooling the colder reservoir $L$ using work $W$, is given by $\eta_r = (-q_{tot}^L)/W$. The corresponding Carnot values are $\eta_r^C = T_L/(T_R - T_L)$, and $\eta_E^C = (T_R - T_L)/T_R$, respectively. Calculated efficiencies, normalized to the Carnot values, for the device in Fig. 2(a,b) are shown in Fig. 3(a). The line along which the condition $\varepsilon_{cs}(V_0) = \varepsilon_S(V_0)$ is fulfilled (and where, by symmetry, $\varepsilon_{cs}(-V_0) = \varepsilon_S(-V_0)$) is visible as a ridge of high normalized efficiency values. To the left of this ridge, where $\varepsilon_{cs}(V_0) < \varepsilon_S(V_0)$, the ratchet operates as a refrigerator: during each half cycle of rocking, the electron flow follows the electrochemical potential gradient, and heat flows against the thermal gradient (Peltier effect). Therefore, the normalized efficiency of a refrigerator, $\eta_r/\eta_r^C$, is shown on the plot for this range. For $\varepsilon_{cs}(V_0) > \varepsilon_S(V_0)$, the heat flow in each half cycle follows the thermal gradient, while electrons flow against the potential gradient. The ratchet does work against the battery, using thermal energy provided by the warm reservoir $R$ (Seebeck effect). Therefore, $\eta_E/\eta_E^C$ is plotted.

Along the line $\varepsilon_{cs}(V_0) = \varepsilon_S(V_0)$ the efficiency coefficients approach their corresponding Carnot values. To show this analytically, we note that after several $q_{R/L}(+V_0)$. Assuming that the width of the 'energy filter', $2\delta$, is much smaller than the energy scales $kT_{R/L}$, over which the Fermi-Dirac distributions vary, one can approximate algebraic steps and using the symmetry of $t(\varepsilon,\pm V_0)$ one can simplify $q_{R/L}^{net}(V_0)$ to

$$q_{R/L}^{net}(V_0) \approx \pm \delta\, t_0 \frac{2t}{h} [\varepsilon_{res}(+V_0) - \mu_{R/L}]\left[ f_L\big([\varepsilon_{res}(+V_0) - \mu_{R/L}]\big) - f_R\big([\varepsilon_{res}(+V_0) + V_0]\big)\right]. \qquad (5)$$

In the quasistatic limit $\delta \to 0$ (where electron flow is quenched) one obtains the efficiency parameters $\eta_E^{\delta \to 0} = 2eV_0/(2\phi + eV_0)$, and $\eta_F^{\delta \to 0} = (2\phi - e\phi z)/2eV_0$, where we used the definition of $\varepsilon_{res}$, and the relation $[\mu_L - \mu_R(+V_0)] = eV_0$. For $\varepsilon_{res}(V_0) = \varepsilon_S(V_0)$ (Eq. (2)), one then recovers the respective Carnot efficiencies. As for any reversible heat engine, the power output goes to zero as reversibility is approached. This is apparent in Fig. 3(c) where the power of the device exhibits a valley along the line where $\varepsilon_{res}(V_0) = \varepsilon_S(V_0)$.

In Fig. 3(b) we compare the exact calculations of $\eta_F(\phi)$ (bolder lines) and $\eta_E(\phi)$ (thinner lines) with $\eta_F^{\delta \to 0}(\phi)$ and $\eta_E^{\delta \to 0}(\phi)$, respectively. As $\delta$ approaches $\varepsilon_S$, the exact calculations for finite $\delta$ increasingly deviate from the values obtained in the limit $\delta \to 0$ (Eq. 5). This is because, when the filter resonance is situated within a few $\delta$ around $\varepsilon_S(V_0)$, refrigeration and heat pumping mix within the transmission range and counteract one another. Consequently, as $\varepsilon_{res}(V_0)$ approaches $\varepsilon_S(V_0)$ from above (decreasing $\phi$), the work output of the heat engine, $W$, eventually becomes negative, and as $\varepsilon_{res}(V_0)$ approaches $\varepsilon_S(V_0)$ from below (increasing $\phi$), the cooling power $-q_L^{net}$ of the refrigerator turns into a heating power. This is also apparent in Fig. 3(d) where $W$ and $-q_L^{net}$ are shown for $\varepsilon_{res}(V_0) < \varepsilon_S(V_0)$ and $\varepsilon_{res}(V_0) > \varepsilon_S(V_0)$, respectively. Maximum power becomes smaller with decreasing $\delta$ because fewer electrons per time unit contribute to the current (Eq. (5)). Note that for $\phi < 0.5 \text{ eV}_0$, Joule heating exceeds the cooling power and refrigerator action is not possible in a rocked device as shown in Fig. 2(a, b).

The analysis presented here for the case of an adiabatically rocked ratchet device can easily be mapped onto the case of a static, periodic ratchet [3]. To do so, the two types of energy filters in Figs. 2(a) and 2(b) would be periodically sequenced between electron reservoirs of alternating temperatures, creating an asymmetric ratchet arrangement as indicated in Fig. 2(c). A macroscopic DC electric field represents an external force against which the heat engine does work. It is straightforward to see that Carnot efficiency can be obtained in principle, if the position of ideal energy filters between heat baths is tuned according to Eq. (2) such that only reversible electron flow is possible.

Edwards et al. proposed the use of resonant tunneling through single electron states in a quantum dot for heat pumping purposes [12, 13]. Energy selectivity has been observed in semiconductor quantum dots, using gates to control position $\varepsilon_{res}$ and width $\delta$ of the transmission resonances, which were confirmed to be Lorentzian with a width determined only by the lifetime of the energy states [14]. To realize one cold-hot-cold section of the device in Fig. 2(c) one could elevate the temperature $T_H$ of a small 2D electron gas region connected via quantum dots to two electron reservoirs at the lower, ambient temperature. Adjustment of $T_H$ is feasible using a heat current and thermopower measurements [15]. While controlling the chemical potential in each reservoir by bias voltages, the output power would be determined from the electric current between reservoirs. To determine the heat engine's efficiency, the heat input, given by the input electric heating power minus heat losses to the host crystal, would be measured in calibration experiments [15]. Given experimental demonstration of all the components [14, 15], such a measurement is feasible in principle, although experimentally challenging.

We note that a different type of reversible heat engine, that operates in the absence of a gradient in chemical potential, is a maser system that receives a photon of one energy from a hot reservoir and deposits a photon of lower energy in a cold reservoir, using the difference as work [16]. That strategy is suitable for photons, but not for particles with mass and charge as considered here.

*Acknowledgements.* This work was supported by the Australian Research Council and by the University of Oregon. The authors thank R. D. Astumian, P.-E. Lindelof, J. M. R. Parrondo and P. Reimann for valuable discussions.

## Figure captions

**Fig. 1:** Electron transfer between reservoirs: (a) in the presence of a difference in electrochemical potentials, (b) in the presence of a temperature difference, and (c) in the presence of both. (d) The Fermi-Dirac distributions in the energy range around $\mu_0 = 0.5(\mu_L + \mu_R)$ for $T_R = 2$ K, $T_L = 0.5$ K, and $V = 0.1$ mV.

**Fig. 2:** (a, b) A hypothetical non-linear rectifier. The energy band structure of the device indicates that of an asymmetric quantum dot forming a resonant tunneling structure, with an energy level position that depends on the voltage. Any higher resonant levels are assumed to be out of the reach of thermally excited electrons, $T_R > T_L$, is assumed. (c) A Brownian heat engine consisting of a periodic, static ratchet potential. Particle flow is against a potential gradient of $\Delta E$ per period. With the position of ideal energy filters (indicated as horizontal lines between warm and cold reservoirs) tuned according to Eq. (2), this ratchet works reversibly.

**Fig. 3:** (a) Efficiency (normalized to the Carnot value) of the model device in Fig. 2(a, b) for $T_R = 1$ K $\approx eV_0 = 0.1$ meV, $r_0 = 1$, and $\delta = 10^{-4}$ meV (lifetime $\tau \approx 3$ ns), as a function of the position of the resonant level, $\phi$, and of $T_L$ (see main text for further details). (b) Normalized efficiency for $T_R = 1$ K, $T_L = 0.5$ K, and $eV_0 = 0.1$ meV. Bold lines (left hand half) are for a refrigerator, thinner lines (right hand half) are efficiencies of a heat engine. The full line is for $\delta \rightarrow 0$ (Eq. (5)). The dashed lines are calculated for, from bottom to top, $\delta = 10^{-2}$, $10^{-3}$, and $10^{-4}$ meV. Data are shown only where the efficiencies are defined. (c) and (d) show the power of the device in Fig. 2(a, b) for the same parameters as used in (a) and (b), respectively. To put the numerical values into context, a cooling power of $10^7$ meV/s corresponds to the heat leaked via phonons to a 2DEG with temperature 0.2 K and area 20 $\mu$m$^2$ in a crystal externally cooled to 0.3 K by a He$^3$ system (estimate based on Eq. 3.1 in [13]).

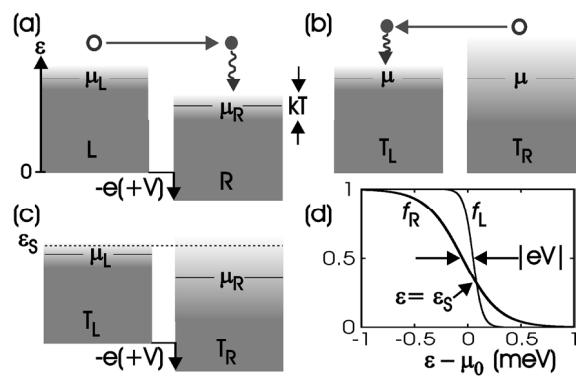

(a)

ε

$\mu_L$

L

0

-e(+V)

$\mu_R$

R

kT

(b)

$\mu$

$T_L$

$\mu$

$T_R$

(c)

$\varepsilon_S$

$\mu_L$

$T_L$

-e(+V)

$\mu_R$

$T_R$

(d)

$f_R$   $f_L$

0.5

0

|eV|

$\varepsilon = \varepsilon_S$

-1   -0.5   0   0.5   1

$\varepsilon - \mu_0$ (meV)

PRL
Humphrey et. al.
Fig. 1

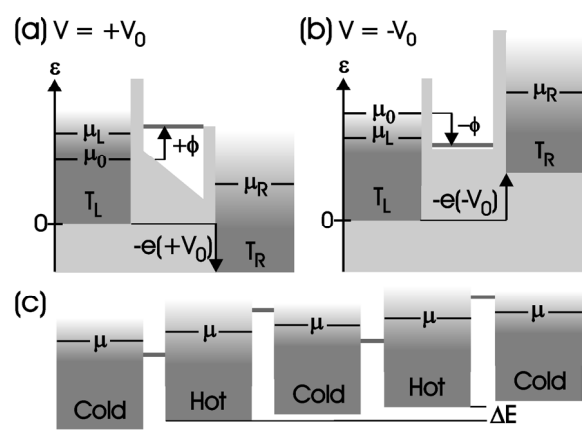

(a) V = +V₀  (b) V = -V₀  (c)

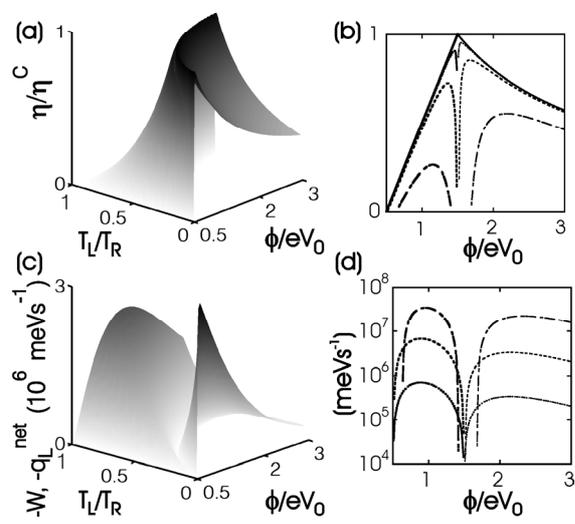